\documentclass[aps,prd,showpacs,amssymb,floatfix,preprintnumbers]{revtex4}
\pdfoutput=1
\newcommand{\be}{\begin{equation}}
\newcommand{\ee}{\end{equation}}
\newcommand{\beq}{\begin{eqnarray}}
\newcommand{\eeq}{\end{eqnarray}}
\usepackage{bm}% bold math
\usepackage{graphicx}
\usepackage{epstopdf}
\usepackage{epsfig}
\usepackage[utf8]{inputenc}
\usepackage{amsmath}
\usepackage{amsfonts,amssymb}
\usepackage{hhline}
\usepackage{multirow}

\newcommand{\dslash}{\not{\hbox{\kern-2pt $\partial$}}}

\preprint{CP3-Origins-2015-053 DNRF90}
\preprint{DIAS-2015-53}
\preprint{CERN-PH-TH-2015-312}

\begin{document}
\title{Large volumes and spectroscopy of walking theories}
\author{L.~Del~Debbio$^{a}$, B.~Lucini$^b$, A.~Patella$^{c,d}$,
  C.~Pica$^{e}$ and A.~Rago$^{d}$} 
\affiliation{$^a$ Higgs Centre for Theoretical Physics, School
  of Physics \& Astronomy, University of Edinburgh, Edinburgh EH9
  3JZ, UK} 
\affiliation{$^b$ College of Science, Swansea
  University, Singleton Park, Swansea SA2 8PP, UK} 
\affiliation{$^c$
  CERN-TH , CH-1211 Geneva, Switzerland}
\affiliation{$^d$
  Centre for Mathematical Sciences, Plymouth University, Plymouth, PL4 8AA, UK}
\affiliation{$^e$
CP$^3$-Origins \& Danish IAS, Campusvej 55, DK-5230 Odense M, Denmark}
%\date{\today}%
\begin{abstract}
  A detailed investigation of finite size effects is performed for
  SU(2) gauge theory with two fermions in the adjoint representation,
  which previous lattice studies have shown to be inside the conformal
  window. The system is investigated with different spatial and
  temporal boundary conditions on lattices of various spatial and
  temporal extensions, for two values of the bare fermion mass
  representing a {\em heavy} and {\em light} fermion regime.  Our
  study shows that the infinite volume limit of masses and decay
  constants in the mesonic sector is reached only when the mass of the
  pseudoscalar particle $M_\mathrm{PS}$ and the spatial lattice size
  $L$ satisfy the relation $L M_\mathrm{PS} \ge 15$. This bound, which
  is at least a factor of three higher than what observed in QCD, is a
  likely consequence of the different spectral signatures of the two
  theories, with the scalar isosinglet ($0^{++}$ glueball) being the
  lightest particle in our model. In addition to stressing the
  importance of simulating large lattice sizes, our analysis
  emphasises the need to understand quantitatively the {\em full}
  spectrum of the theory rather than just the spectrum in the mesonic
  isotriplet sector.  While for the lightest fermion measuring masses
  from gluonic operators proves to be still challenging, reliable
  results for glueball states are obtained at the largest fermion mass
  and, in the mesonic sector, for both fermion masses.  As a byproduct
  of our investigation, we perform a finite size scaling of the
  pseudoscalar mass and decay constant.  The data presented in this
  work support the conformal behaviour of this theory with an
  anomalous dimension $\gamma_* \simeq 0.37$.
\end{abstract}
\maketitle

\section{Introduction}
\label{sec:introduction}

The recent discovery of the Higgs boson is the first step towards new
physics beyond the Standard Model. As the properties of this new state
are investigated in more detail, the constraints on models of new
physics will become more stringent. It has been suggested in a long
line of works stemming from
Refs.~\cite{Weinberg:1975gm,Susskind:1978ms,Eichten:1979ah} that at a
more fundamental level the Higgs sector and the electroweak symmetry
breaking mechanism of the Standard Model could be explained
dynamically in terms of a novel strong interaction. In the light of
recent developments, the study of scenarios of strongly interacting
gauge theories beyond the Standard Model is playing an increasingly
central role in both theoretical modeling and experimental
searches. For instance, a preliminary observation of an excess in the
di-photon decay channel has been recently reported by both ATLAS and
CMS~\cite{atlas:2015dig,CMS:2015dxe}. A strong interaction beyond the Standard Model can
provide a natural interpretation of this excess. In the framework of a
novel strong force, the Higgs particle would be a composite state of
the new interaction. The phenomenology of such a composite state is
usually studied using effective theory descriptions, which rely on
general (and possibly few) assumptions about the new dynamics. 
Nonperturbative effects due to the novel
strong force are encoded in the low energy constants of the effective
lagrangians.  A naive guess for the candidate theory that describes
this new force could be a gauge theory with matter coupled to the
electroweak sector of the Standard Model. It has been known for a long
time that the simplest model of this type, a rescaled version of QCD,
would fail to satisfy the constraints from precision electroweak
data. A more quantitative understanding of the nonperturbative
dynamics of gauge theories is required in order to develop and test
models that go beyond a naive rescaling of QCD. Depending on the gauge
and matter content, these theories can be rather different from
QCD. Computing from first principles their spectrum, and other
low-energy constants relevant for phenomenology, would provide a most
useful input for phenomenological studies. Unfortunately these
quantities are notoriously difficult to compute for realistic
candidate theories.

Lattice simulations with dynamical fermions have reached a
stage where light fermion masses can be simulated effectively
on large lattices using existing hardware. Therefore numerical methods
provide an effective tool to investigate the phase diagram of gauge
theories from first principles, and indeed several extensive studies
have been performed in recent years, exploring the theory space by
varying the number of colors and flavors, as well as the color
representation of the fermionic matter.

Note that, in order to satisfy the constraints from electroweak
precision data, and in particular to build a realistic flavor sector,
the new theory is required to be near-conformal, i.e. to have a
multi-scale dynamics, characterized by a clear separation between an
IR and an UV scale. It is often advocated that such a separation is
obtained by considering theories that are just outside the conformal
window, or deformations of theories that have an infrared fixed point
(IRFP), hence the ``near-conformal'' denomination. A large energy
range where the RG evolution of the couplings is very slow ({\em
  walking}) is expected to appear, which leads to an approximate scale
invariance at large distances. The low-energy dynamics is then
characterized by the anomalous dimensions of the relevant operators at
the IRFP, and large anomalous dimensions are typically required in
order to construct phenomenologically viable
models~\cite{Yamawaki:1985zg,Appelquist:1987fc,Hong:2004td,Dietrich:2005jn,Luty:2004ye,Luty:2008vs}.
Lattice simulations yield quantitative results in strongly interacting
theories from first principles. Holographic theories are an
alternative way to perform explicit calculations using the
perturbative dual description of the strong dynamics; several efforts
to understand near-conformality in this framework have appeared over
the years (see for
instance~\cite{Contino:2003ve,Agashe:2004rs,Hong:2006si,Hirn:2008tc,Nunez:2008wi,Elander:2009pk,Anguelova:2010qh,Elander:2013jqa,Alho:2013dka,Hoyos:2013gma,Faedo:2013ota,Evans:2014nfa,Erdmenger:2014fxa}).

The main goal of our current numerical studies is the unambiguous
identification of an IRFP, and the determination of the anomalous
dimensions that characterize the dynamics. This is a difficult
problem, since the systematic errors that affect lattice simulations
need to be kept under control in order to highlight the interesting
conformal behaviour. First studies, which have been useful in the
early stages of the numerical investigations, are affected by large
systematics, which can obscure the physical features of these
theories. Different approaches to conformality on the lattice have
been suggested and tested over the
years~\cite{Appelquist:2007hu,Hietanen:2008mr,Hietanen:2009az,Karavirta:2011zg,Shamir:2008pb,Deuzeman:2009mh,DelDebbio:2010hx,DelDebbio:2010ze,DelDebbio:2010jy,DelDebbio:2013qta,Bursa:2009we,Catterall:2011zf,DeGrand:2011cu,Patella:2012da,Ishikawa:2013wf,Ishikawa:2015iwa}.
They include a variety of methods that range from the scaling of the
spectrum to the explicit computation of the running of the couplings.
Despite several large scale projects devoted to these
problems~\cite{Aoki:2012eq,Aoki:2013zsa,Aoki:2014oha,Fodor:2011tu,Fodor:2012ty,Appelquist:2011dp,Appelquist:2014zsa},
there is still some controversy in the literature, showing that great
care must be exercised in the analysis of the lattice data. 

In this work we focus on SU(2) gauge theory coupled to $n_f=2$ flavors
of Dirac fermions in the adjoint representation of the gauge group.
This model, introduced as a phenomenological extension of the Standard
Model under the name of Minimal Walking Technicolor
(MWT)~\cite{Sannino:2004qp}, is believed to have an IR fixed
point~\cite{DelDebbio:2010hu,DelDebbio:2010hx,Bursa:2009we,DelDebbio:2009fd,Karavirta:2011mv,Hietanen:2008mr,Hietanen:2009az,Rantaharju:2015yva}.
We would like to find evidence for the conformal behaviour from a
study of the spectrum of this theory. The scaling of the spectrum, and
the corrections to scaling, allow in principle a determination of the
anomalous dimensions of the theory at the IR fixed point, see e.g.
Refs.~\cite{DeGrand:2009hu,DelDebbio:2010ze,DelDebbio:2010jy,DelDebbio:2013qta}. A
precise determination of the spectrum would also allow a test of the
Feynman-Hellmann relations that were proposed in
Ref.~\cite{DelDebbio:2013sta}. The systematic errors on the spectrum
due to the finite temporal and spatial extents of the lattice for MWT
have been analysed in a previous publication~\cite{Bursa:2011ru}:
systematic errors of approximately 10\% are common on lattices such
that $L M_\mathrm{PS} < 10$, where $M_\mathrm{PS}$ indicates the mass of
the lightest pseudoscalar state in the spectrum. We have therefore
embarked in large volume simulations of the theory, in order to
provide results for the spectrum in a regime where systematic errors
are below 1\%. The simulations have been performed using the {\tt
  HiRep} code developed in Ref.~\cite{DelDebbio:2008zf}. A detailed
description of the methodology, and of the lattices simulated in this
work is reported in Sect.~\ref{sec:method}.

The new volumes simulated are large enough to avoid contamination from
excited states effects, and to allow us to extrapolate the data for
the spectrum to the infinite volume limit for two values of the
fermion mass. These are the first results for the spectrum of the MWT
that can be extrapolated to the thermodynamical limit with an
uncertainty at the percent level. The results for the full spectrum,
including glueball states and the string tension, are discussed in
Sect.~\ref{sec:results}.

\section{Methodology}
\label{sec:method}

We have simulated MWT using the RHMC algorithm described in
Ref.~\cite{DelDebbio:2008zf}. In order to identify, and control, the
systematic effects due to the finite size of the lattices, we have
simulated the theory on a series of lattices, increasing both the
temporal and the spatial extent of the system. The new runs are listed
in Tab.~\ref{tab:listruns}. All simulations have been performed at
fixed lattice bare coupling $\beta=2.25$, and for two values of the
fermion bare mass $am_0=-1.05,-1.15$. As a further tool to investigate
finite volume effects, we have compared the spectrum obtained from
simulations with the usual periodic boundary conditions, to the one
obtained with twisted boundary conditions, as defined in
Ref.~\cite{'tHooft:1979uj}, and the one obtained with open boundary
conditions, as described in Ref.~\cite{Luscher:2012av}. The detailed
implementation of the twisted and open boundary conditions is
described below. In the infinite volume limit, results should be
independent of the boundary conditions, and therefore we can use the
dependence on the boundary conditions to monitor whether or not the
theory has reached the large volume asymptotic behaviour. As discussed
in Ref.~\cite{DelDebbio:2010hx}, the smaller lattices A10 and A11 are
in a regime where the distribution of the values of the Polyakov loop
displays a double-peak structure. This regime where centre symmetry is
broken is usually not suitable for extracting the glueball
spectrum~\cite{DelDebbio:2010hx}. The statistics for all lattices is
targeted to a precise measurement of the mesonic spectrum, which is
less demanding than the glueball analysis. The temporal size of the
lattice has been increased to 64 in order to exploit the long plateau
in the mesonic correlators. The lattices C5, D4, F1, and G1 are in the
regime in which the Polyakov loop has a single peak at zero and is
symmetric around it, and the statistics accumulated (of order 2000
configurations) is suitable for the study of the glueball
spectrum. The large number of trajectories generated on the B2 lattice
is needed because of the larger autocorrelation time observed for this
particular choice of parameters. Note that the trajectory length of
the HMC for this lattice is $t_\mathrm{traj}=1$, shorter than the one
for successive runs. The runs on the twisted lattices are
designed to check the finite volume effects in the mesonic
spectrum. The statistics is chosen accordingly to yield a good signal
in those channels.

\begin{table}[ht]
  \centering
  \begin{tabular}{ccccccccc}
    \hline
    lattice & V & $-am_0$ & $N_\mathrm{traj}$ & $t_\mathrm{traj}$ & $\langle P\rangle$ & $\tau_P$ & $\lambda$ & $\tau_\lambda$ \\
    \hline
    % apatelt4@thqcdgw02
    % /data/apatelt4/Runs/64x8x8x8nc2rADJnf2b2.250000m1.150000
    A10 & $64\times 8^3$  & 1.15 & 810 & 3   & 0.66536(22) & 3.6(1.2) & 0.2005(58) & 1.42(31) \\
    % apatelt4@thqcdgw02
    % /data/apatelt4/Runs/64x12x12x12nc2rADJnf2b2.250000m1.150000
    A11 & $64\times 12^3$ & 1.15 & 530 & 1.5 & 0.66601(15) & 1.93(59) & 0.2054(40) & 1.54(43) \\
    % bgdbl@bluecr-bg-fen
    % /gpfs/bgdbl/Data/SU2/Adj/V64x16/run1_B2.25_M-1.15_V64x16_APT
    C5 & $64\times 16^3$ & 1.15 & 1500 & 1.5 & 0.665992(61) & 2.32(46) & 0.2116(16) & 2.38(48) \\
    % bgdbl@bluecr-bg-fen
    % /gpfs/bgdbl/Data/SU2/Adj/V64x24/run1_B2.25_M-1.15_V64x24_APT
    D4 & $64\times 24^3$ & 1.15 & 2387 & 1.5 & 0.665927(26) & 3.92(79) & 0.21478(70) & 1.70(24) \\
    % bgdbl@bluecr-bg-fen
    % /gpfs/bgdbl/Data/SU2/Adj/V64x32/run1_B2.25_M-1.15_V64x32_APT
    F1 & $64\times 32^3$ & 1.15 & 2541 & 1.5 & 0.665946(30) & 3.37(62) & 0.2115(12) & 1.06(12) \\
    G1 & $80\times 48^3$ & 1.15 & 2200 & 1.5 & 0.665943(17) & 5.1(1.2) & 0.2237(17) & 0.637(58) \\
    G2 & $80\times 48^3$ & 1.15 & 3436 & 1.5 & 0.665933(18) & 5.6(1.8) & 0.2235(14) & 0.639(26) \\
    \hline
    % old
    B2 & $24\times 12^3$ & 1.05 & 7819 & 1 & 0.647633(70) & 6.79(99) & 1.4936(51) & 5.80(78) \\
    % pybiagio@pygrid
    % /mnt/disk01/8/swansea/pybiagio/BSM/Data/SU2/Adj/V64x16/run4_B2.25_M-1.05_V64x16
    C6 & $64\times 16^3$ & 1.05 & 2648 & 1.5 & 0.647645(48) & 4.63(96) & 1.4389(36) & 1.26(14) \\
    % bgdbl@bluecr-bg-fen
    % /gpfs/bgdbl/Data/SU2/Adj/V64x24/run1_B2.25_M-1.05_V64x24_APT
    D5 & $64\times 24^3$ & 1.05 & 4000 & 1.5 & 0.647695(37) & 3.56(53) & 1.3906(45) & 0.722(54) \\
    % bgdbl@bluecr-bg-fen
    % /gpfs/bgdbl/Data/SU2/Adj/V48x32/run1_B2.25_M-1.05_V48x32_APT
    F2 & $48\times 32^3$ & 1.05 & 3590 & 1.5 & 0.647680(30) & 4.28(74) & 1.3708(49) & 0.632(45) \\
    \hline
    \hline
    % apatelt4@thqcdgw02
    % /data/apatelt4/Runs/64x8x8x8nc2rADJnf2b2.250000m1.150000_TBC
    TWA1 & $64\times 8^3$  & 1.15 & 565 & 1.5 & 0.66665(22) & 2.8(1.0) & 0.5557(96) & 0.85(18) \\
    % apatelt4@thqcdgw02
    % /data/apatelt4/Runs/64x12x12x12nc2rADJnf2b2.250000m1.150000_TBC/run1
    TWB1 & $64\times 12^3$ & 1.15 & 741 & 1.5 & 0.66590(11) & 2.96(96) & 0.2709(48) & 1.93(50) \\
    % apatelt4@thqcdgw02
    % /data/apatelt4/Runs/64x16x16x16nc2rADJnf2b2.250000m1.150000_TBC/run1
    TWC1 & $64\times 16^3$ & 1.15 & 1162 & 1.5 & 0.665990(61) & 2.91(73) & 0.2484(17) & 6.2(2.2) \\
    TWD1 & $64\times 24^3$ & 1.15 & 2701 & 1.5 & 0.665912(35) & 4.63(95) & 0.21840(88) & 2.43(37) \\
    \hline
    \hline
    OG1 & $80\times 48^3$ & 1.15 & 1248 & 1.5 & 0.643620(16) & 1.12(2) & 0.22607(58) & 6.12(2) \\
    \hline
    \hline
  \end{tabular}
  \caption{List of lattices used in this study. The nomenclature for
    the lattices follows our earlier conventions. The lattice denoted
    by TW are the ones with twisted boundary conditions.  The lattices
    denoted by O are the ones with open boundary conditions in one of
    the four Euclidean directions. The new runs
    are performed at two different values of the bare mass
    $am_0$. $N_\mathrm{traj}$ is the number of trajectories (of length
    $t_\mathrm{traj}$) that we have generated. The value of the plaquette
    $\langle P\rangle$ and of the lowest eigenvalue of the Dirac 
    operator $\lambda$ are also reported together with their respective
    integrated autocorrelation times $\tau_P$ and $\tau_{\lambda}$. All
    runs are performed at $\beta=2.25$.}
  \label{tab:listruns}
\end{table}

\subsection{Observables}
\label{sec:observables}

All observables discussed in this paper are obtained from the expectation values of field
correlators, using techniques that are standard in lattice simulations.
Mesonic observables are extracted from two-point functions of fermion bilinears:
\begin{equation}
  \label{eq:mes2pt}
  f_{\Gamma\Gamma^\prime}(t) = \sum_{\vec{x}}\, \langle \Phi_\Gamma(\vec{x},t)^\dagger
  \Phi_{\Gamma^\prime}(\vec{0},0)\rangle\, , 
\end{equation}
where 
\begin{equation}
  \label{eq:fbil}
  \Phi_\Gamma(\vec{x},t) = \bar\psi_1(\vec{x},t) \Gamma \psi_2(\vec{x},t)\, .
\end{equation}
Note that we always consider non-singlet flavor states, as indicated
by the indices $1,2$ that appear in the definition of the fermion
bilinear. The matrices $\Gamma$ and $\Gamma^\prime$ act in spin space,
and determine the quantum numbers of the states that contribute to the
correlator in Eq.~(\ref{eq:mes2pt}). The channels studied here are
obtained by choosing
$\Gamma=\Gamma^\prime=1,\gamma_0\gamma_5\gamma_k,\gamma_5\gamma_k,
\gamma_0\gamma_k, \gamma_k,\gamma_5$.  The details of the analysis
used to extract the PCAC mass, the hadron masses, and the decay
constants are provided in the Appendix of
Ref.~\cite{Bursa:2011ru}. Gluonic observables are obtained using the
techniques described in Ref.~\cite{Lucini:2010nv}.

\subsection{Contamination from excited states}
\label{sec:finite-temp-effects}

\begin{figure}[ht] 
  \includegraphics*[scale=0.35]{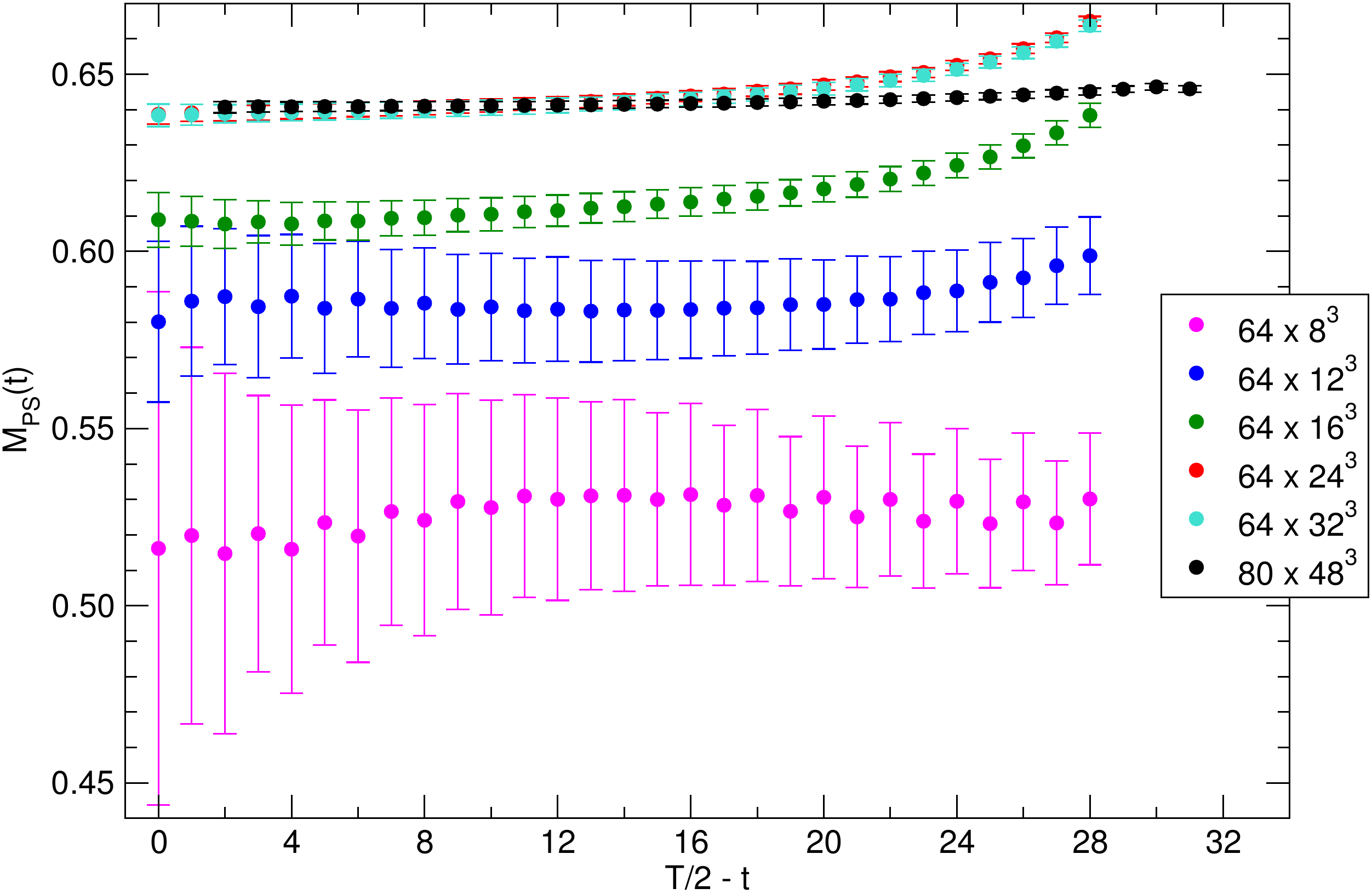} 
  \caption{Plateaux for the effective mass in the PS channel at
    $\beta=2.25$ and $-a m_0 = 1.15$ for periodic boundary
    conditions. The figure shows that mass plateaux as a function of
    $t$ are clearly identified for the current range of parameters,
    and that there is no contamination from higher states in the
    spectrum for the lattices considered here, i.e. with
    $L/a \geq 64$.}
\label{fig:15plateaux}
\end{figure}

We had noticed in our previous simulations that lattices with a
temporal extent $T/a=16, 24$ are not long enough to identify
unambiguously the onset of the asymptotic behaviour of the field
correlators~\cite{Bursa:2011ru}. As discussed below, this is not
unexpected if the theory has an IR fixed point. Larger time extensions
have been used in this study in order to be able to identify long
plateaux in the effective mass plots, and to fit the large time
behaviour in a regime where contaminations from higher states in the
spectrum have died out. A typical set of plateaux for the mass of the
pseudoscalar meson is reported in Fig.~\ref{fig:15plateaux}. Lattices with temporal extent $T/a\geq 64$
exhibit long plateaux for the effective mass, which can be fitted to a
constant over a large range in $t$. As a consequence, we decided not
to use smeared sources for this analysis, since the introduction of
smeared sources increases the auto-correlation time of the spectral
observables.

\subsection{Twisted boundary conditions}
\label{sec:twist-bound-cond}

In order to investigate in more detail the systematic errors introduced
by the finite extent of the spatial volume, we have performed a series
of simulations with twisted boundary conditions. Twisted boundary
conditions, as introduced by 't~Hooft~\cite{'tHooft:1979uj}, can be
implemented in a theory with matter in the adjoint representation. We
consider here twisted boundary conditions in the spatial directions,
which introduce a magnetic flux in the system. In the confined phase,
the effect of these boundary conditions should vanish as the spatial
size of the box is taken to infinity. Any difference in the spectrum
as the boundary conditions are varied could signal sizeable finite volume
effects. 

Let us summarize the implementation of the boundary conditions used in
this work.  Twisted boundary conditions are imposed for both the gauge
and the fermion fields:
\begin{gather}
A_\mu(x_0, \bm{x} + L \bm{e}_k) = \Omega_k A_\mu(x_0,\bm{x}) \Omega_k^\dag \ , \\
\psi(x_0, \bm{x} + L \bm{e}_k) = \Omega_k \psi(x_0,\bm{x}) \Omega_k^\dag \ ,
\end{gather}
where $\Omega_k$ are generic elements of the gauge group. For the fields to be single-valued, the following constraint must be satisfied:
\begin{gather}
\Omega_j \Omega_k = e^{\frac{2 \pi i}{N} \epsilon_{jkl} m_l} \Omega_k \Omega_j \ ,
\label{eq:magnetic_flux}
\end{gather}
where $m_l \in \mathbb{Z}$ if the gauge group is SU(N). Twisted boundary conditions introduce
three orthogonal Dirac strings aligned along the $\bm{e}_1$, $\bm{e}_2$ and $\bm{e}_3$ directions,
each one carrying center-magnetic flux equal to $m_1$, $m_2$ and $m_3$ respectively.
Note that physical observables only depend
on the magnetic flux $\bm{m}$, and not on the particular choice of the $\Omega_k$ matrices.

For the simulations presented here, i.e. for the case $N=2$, we choose the maximally symmetric twist:
\begin{gather}
\bm{m} = (1,1,1) \ ,
\end{gather}
which preserves the octahedral isometries (including parity) and charge-conjugation symmetry.

Twisted boundary conditions are implemented on the lattice by
modifying the action. The link variables and fermions are still
periodic, but the gauge part of the action is modified by multiplying
the plaquettes by a center element. For a theory with SU(2) gauge group
\begin{gather}
S_g = -\frac{\beta}{4} \sum_{x,\mu<\nu} z_{\mu\nu}(x) \textrm{tr}\ U_{\mu\nu}(x) \ .
\end{gather}
The contribution to the action of a given plaquette is multiplied by
-1, if the plaquette is pierced by the Dirac string, i.e., with $\mu$,
$\nu$ and $\rho$ the three spatial directions and indicating with zero
the temporal direction: 
\begin{gather}
z_{\mu\nu}(x) =
\begin{cases}
-1 & \textrm{for} \ x_0=x_\rho=1 \\
1 & \textrm{otherwise}
\end{cases} \ \ , 
\end{gather}
for the three possible choices of $\mu < \nu$. The fermion part of the
action remains unchanged, and therefore only the gluonic force is
modified in the molecular dynamics evolution.  

\subsection{Open boundary conditions}
\label{sec:open-bound-cond}

In a setup with periodic boundary conditions in all directions, the
autocorrelation times increase exponentially with the lattice spacing
because of the existence of topological sectors in the continuum
limit. Neumann (open) boundary conditions remove the infinite-energy
barrier between topological sectors.  For the range of lattice
spacings studied in Ref. ~\cite{Luscher:2011kk} in the case of QCD,
the autocorrelation times scale with $a^{-2}$.
% It has been shown in
% Ref. ~\cite{Luscher:2011kk} that the autocorrelation times scale with
% $a^{-2}$ in this case.

Even though our simulations are performed at fixed $\beta$, we have
observed a very poor scaling of the autocorrelation time of the
topological susceptibility towards the chiral limit (which might be a
peculiarity of IR conformality). Simulations with open boundary
conditions have been performed using a modified version of the {\tt
  HiRep} code, showing that indeed the choice of boundary conditions
ameliorate this problem.

We have generated only one lattice with open boundary conditions,
denoted OG1 in Tab.~\ref{tab:listruns}. The parameters for this run
correspond to the ones used for the runs G1 and G2 with periodic
boundary conditions. These are our largest lattices at the lighter
fermions mass. This extra set of simulations has been added to address
specifically the problems that we observe in the glueball spectrum at
the lighter mass. As the glueballs are the lightest states in this
theory, it is important to have a quantitative understanding of the
finite volume effects for these observables. 

As in the case of periodic boundary conditions, spectral quantities
can be extracted from two-point correlators of composite fields with
the approriate quantum numbers. This follows from the observation that
open boundary conditions correspond in Hamiltonian formalism to the
insertion of a pure state with quantum numbers of the vacuum, see
e.g. Refs.~\cite{Bruno:2014lra,Bruno:2014jqa,Luscher:2012av} for a
recent discussion. We notice that in this case the two-point functions
cannot be averaged over time, as translational invariance is lost in
the time direction. The details of our analysis are presented in the
following section.

\section{Results}
\label{sec:results}

\subsection{Heavier fermion mass}
\label{sec:heavier-fermion-mass}

\begin{figure}[ht] 
  \includegraphics*[scale=0.35]{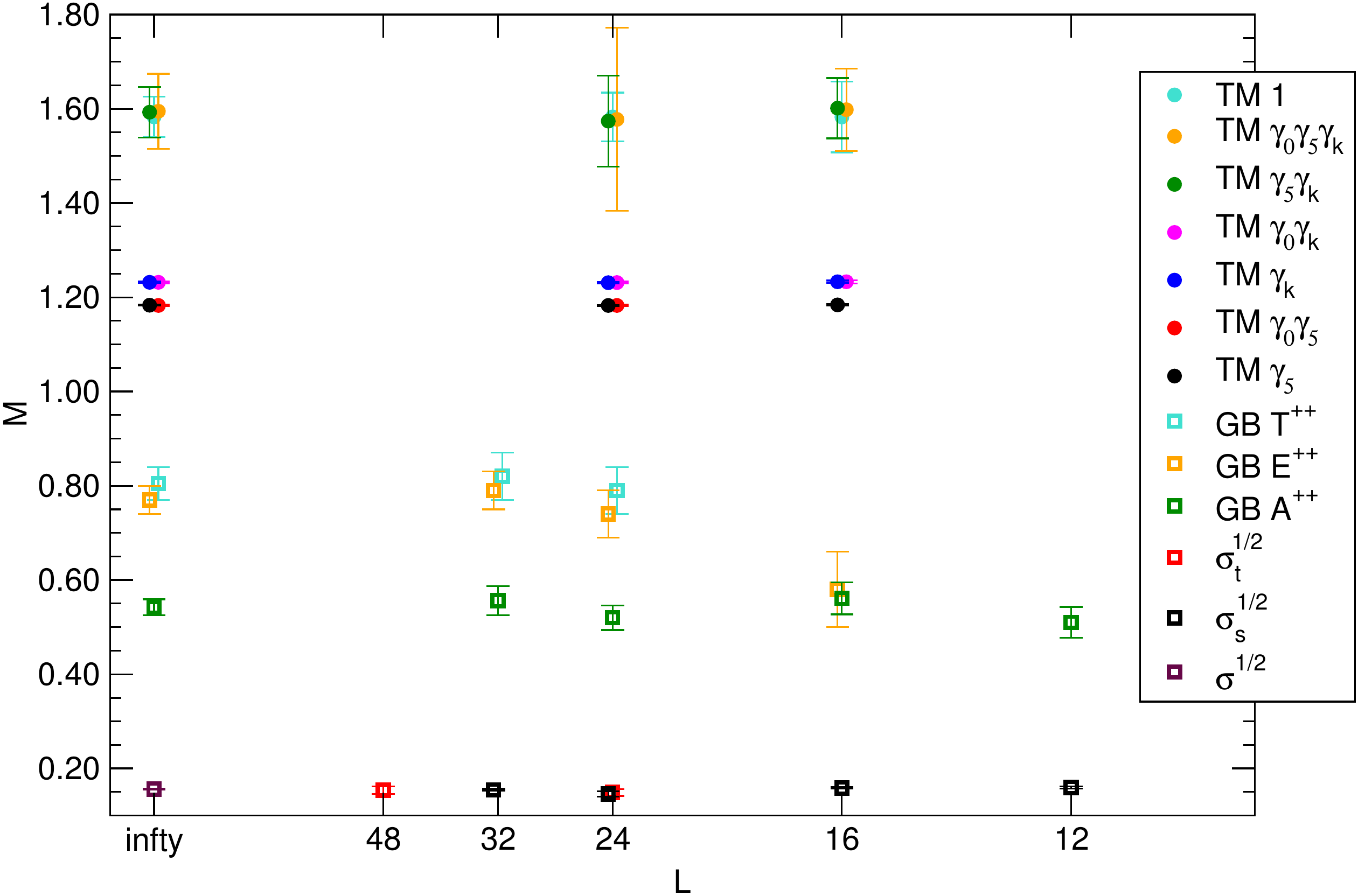} 
  \caption{ Summary of the spectrum for all the states included in
    this analysis at $am_0=-1.05$ with periodic boundary
    conditions. The data are shown as a function of the number of
    sites in the spatial directions $L/a$.  }
\label{fig:05all}
\end{figure}

\begin{figure}[ht] 
  \includegraphics*[scale=0.35]{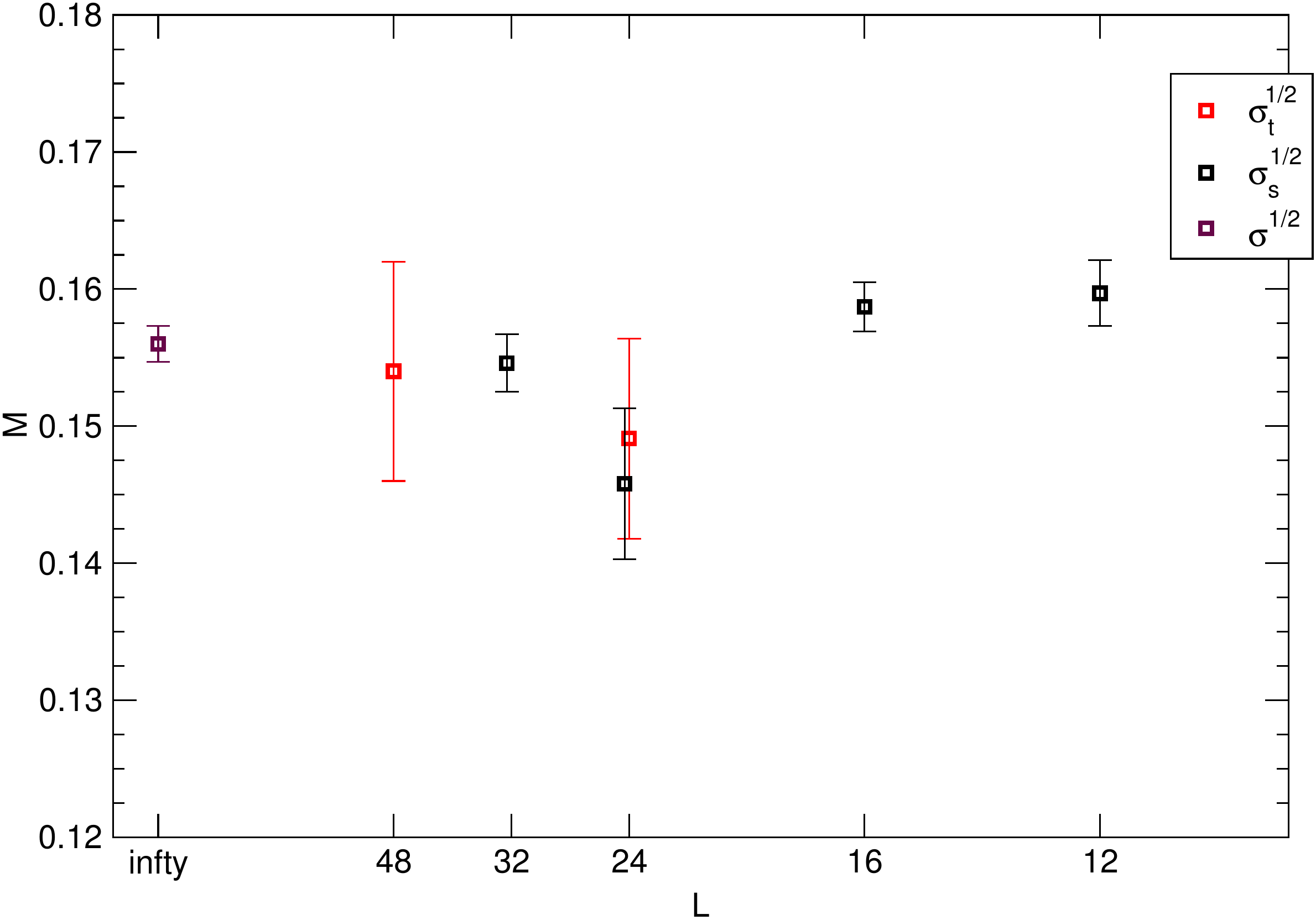} 
  \caption{ String tension extracted from spatial and temporal
    Polyakov loops at $am_0=-1.05$ with periodic boundary
    conditions. The results are shown as a function of the number of
    sites in the spatial directions $L/a$.  }
\label{fig:05sigma}
\end{figure}

The heavier fermion mass in the new set of simulations is $am_0=-1.05$, which corresponds to a
PCAC mass of $am=0.2688(15)$. The full spectrum is shown in Fig.~\ref{fig:05all}. As observed in
previous studies the mesons are heavier than the glueballs, with the smallest scale being set by the
string tension. On the largest volumes, we observe that the lightest
glueball state is in the $A^{++}$ channel (which corresponds to the
continuum $0^{++}$) and the groundstates in the $E^{++}$ and in the
$T^{++}$ channels (which both single out the continuum $2^{++}$) have
the expected degeneracy. It is worth noting that the string tension can be obtained from
correlators of Polyakov loops wrapping either in temporal or in spatial direction, the two
measures are characterised by finite volume effects driven by
the spatial or the temporal size respectively. In
Fig.~\ref{fig:05all} and later in Fig.~\ref{fig:05sigma} we report
both measures as a function of their relevant size.

\begin{figure}[ht] 
  \includegraphics*[scale=0.35]{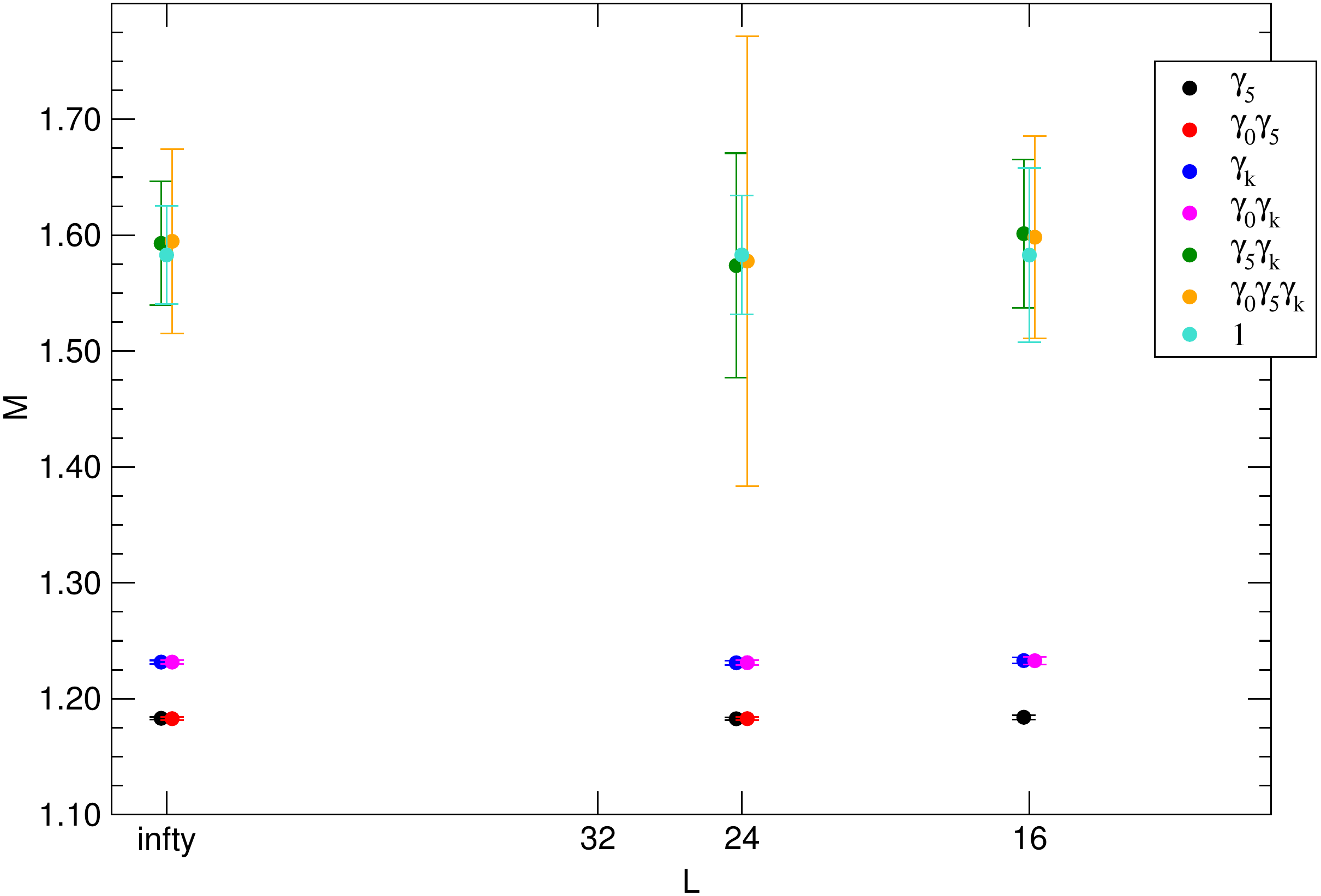} 
  \caption{ Summary of the spectrum of the mesonic states included in
    this analysis at $am_0=-1.05$ with periodic boundary
    conditions. The data are shown as a function of the number of
    sites in the spatial directions $L/a$.  }
\label{fig:05mesons}
\end{figure}

The detailed analysis of the mesonic spectrum is shown in Fig.~\ref{fig:05mesons}. At this value of
the fermion mass, the finite volume effects on the mesonic states are small, and we can confidently
extrapolate to the thermodynamical limit. Different combinations of $\Gamma,\Gamma^\prime$ that
project onto the same physical states yield results that are
compatible within statistical errors. The new simulations confirm the
results that we presented in previous studies~\cite{Bursa:2011ru}.

The results for the string tension are reported in
Fig.~\ref{fig:05sigma}. Simulations on sufficiently
large lattices, $L/a\geq 24$, show a satisfactory agreement between
the spatial and temporal string tensions.

\subsection{Lighter fermion mass}
\label{sec:lighter-fermion-mass}

Let us now consider the lightest mass, where the finite volume effects
are expected to be larger. The mass of the pseudoscalar, reported in
Fig.~\ref{fig:15plateaux}, is affected by sizeable finite volume
corrections on the smaller lattices; these corrections can be as large
as $23 \%$ if $L M_\mathrm{PS} \simeq 4$ and a well visible $5 \%$ for
$L M_\mathrm{PS} \simeq 9.75$.  The fitted value seems to converge for
lattices with $L/a\geq 24$, corresponding to
$L M_\mathrm{PS} \simeq 15.4$.  Moreover, this plot shows that the
effective mass is growing as the spatial size of the lattice is
increased. This is consistent with the observation that, in a theory
where chiral symmetry is not spontaneously broken, the pseudoscalar
mass reaches the infinite volume limit from below, as explained in
Refs.~\cite{Luscher:1985dn,Koma:2004wz,Patella:2011kp}.

The plateaux in the effective mass of the pseudoscalar are reported as
a function of the time separation between the interpolating operators
$t$. They show that a large temporal extent is also required in
order to suppress the corrections from heavier states in the same 
channel. This is expected for a mass deformed conformal theory, since
{\em all} the states in the spectrum should scale to zero following
the same power-law scaling formula. The contribution from excited
states becomes neglibile when: \begin{equation}
  \label{eq:Deltam}
  t \Delta M_\mathrm{PS} \gg 1\, 
\end{equation}
where $\Delta M_\mathrm{PS}$ is the mass difference between the ground state and the first excited
stated in the pseudoscalar channel. For a conformal theory near the chiral limit, $\Delta
M_\mathrm{PS} \propto M_\mathrm{PS}$, so that the exponential suppression of the excited states
becomes really effective only at large temporal separations.

\begin{figure}[ht] 
  \includegraphics*[scale=0.35]{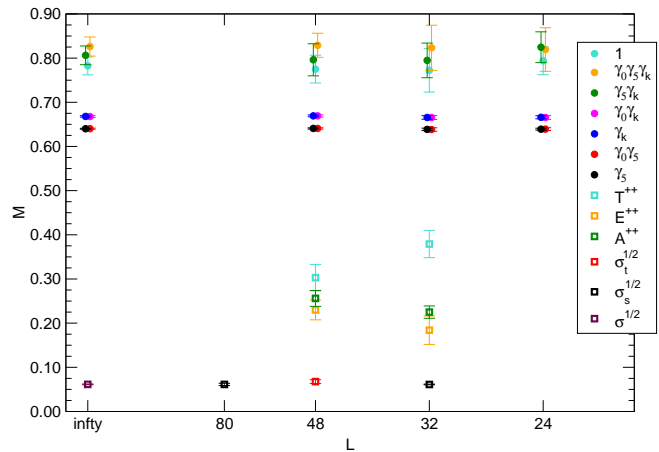} 
  \caption{ Summary of the spectrum for the lattices included in this
    analysis at $am_0=-1.15$ with periodic boundary conditions. The
    data are shown as a function of the number of sites in the spatial
    directions $L/a$.  }
\label{fig:15all}
\end{figure}

The data for the full spectrum including the glueballs is shown in
Fig.~\ref{fig:15all}. Note that in this case the glueball spectrum
shows a lack of degeneracy between the groundstate masses in the
$E^{++}$ and in the $T^{++}$ channels and a near-degeneracy of the
$A^{++}$ channel and of the $E^{++}$, with the latter mass having the
tendency to be lighter. This is at odds with the expected behaviour as
the continuum limit is approached, where the $E^{++}$ and $T^{++}$
channels should become degenerate, and heavier than the scalar state
$A^{++}$. This unexpected behaviour has already been observed
in Ref.~\cite{DelDebbio:2010hx}, where it was ascribed to 
finite size effects arising at the cross-over between the small volume
regime and the regime connected to the thermodynamic limit. This
interpretation is supported by our current study at higher mass, where
hints of those phenomena can be seen on the smallest lattices, but
then disappear when the size of lattices is increased, as shown in
Fig.~\ref{fig:05all}. 

As discussed above, finite-volume effects can be investigated by
comparing results obtained with different boundary conditions; in
particular for the glueball spectrum the comparison between periodic
and open boundary conditions yields interesting results. The spectrum
is obtained from the large time behaviour of correlators far away from
the boundaries, while trying to average over
different times in order to improve the statistical signal. More
precisely, we computed the correlators:
\begin{equation}
  \label{eq:gluecorrOBC}
  f_{12}(t) = \frac{1}{N_\mathrm{pts}} \sum_{n=-\left\lfloor
      N_\mathrm{pts}/2 \right \rfloor +1}^{\left\lfloor
      N_\mathrm{pts}/2 \right\rfloor} \langle \mathcal{O}_1(T/2+n+t/2)
  \mathcal{O}_2(T/2+n-t/2)\rangle\, ,
\end{equation}
with a straightforward modification in the case where $t$ is odd. For
a given time separation $t$ between the interpolating fields
$\mathcal{O}_1$ and $\mathcal{O}_2$, we average over $N_\mathrm{pts}$
time planes, the values of $N_\mathrm{pts}$ are chosen so that the
$T/2+\left\lfloor N_\mathrm{pts}/2 \right \rfloor+t/2$ and
$T/2-\left\lfloor N_\mathrm{pts}/2 \right \rfloor+1-t/2$ are at a
distance of at least 7 lattice spacings from the boundary. 

\begin{figure}[ht]
  \centering
  \includegraphics*[scale=0.45]{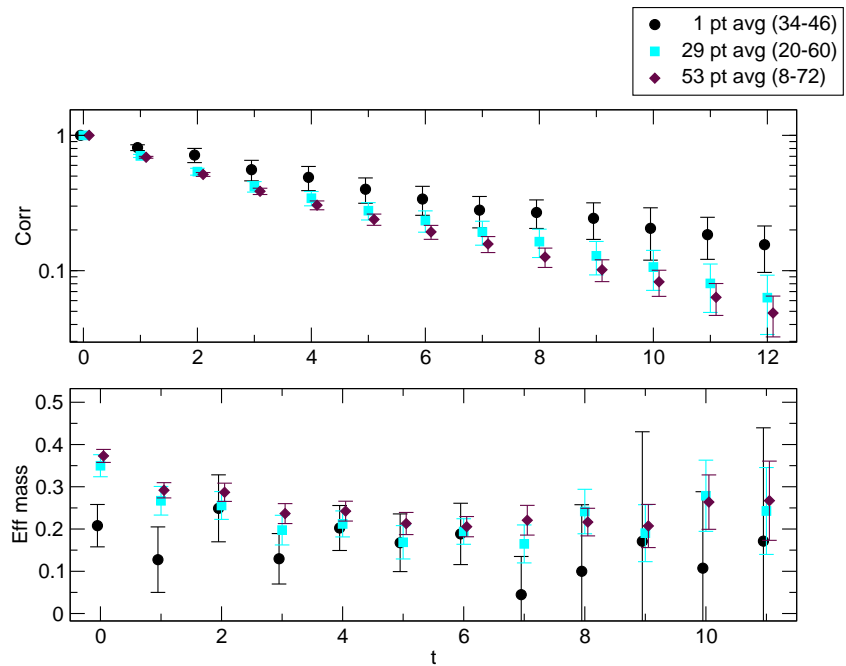} 
  \caption{The correlator $f_{12}(t)$ defined in
    Eq.~(\ref{eq:gluecorrOBC}) (top) and the corresponding effective
    mass (bottom) for the $E^{++}$ glueball as a function of the time
    separation $t$. Simulations are performed with open boundary
    conditions. Different symbols correspond to averages over
    different numbers of time planes, i.e. to choosing $N_\mathrm{pts}=1,29,53$.  }
  \label{fig:oBCEpp}
\end{figure}

The results for the correlator in the $E^{++}$ channel are shown in
Fig.~\ref{fig:oBCEpp}. The averaging over several time
planes improves the determination of the correlators, and the
effective mass plot shows a plateau for time separations $t\geq
4$. Although we analysed several values of $N_\mathrm{pts}$, we
discuss in details the results for $N_\mathrm{pts}=29$. Other
channels, show a similar behaviour for both the correlators and the
effective masses. 

\begin{figure}[h!]
  \centering
  \includegraphics*[scale=0.3]{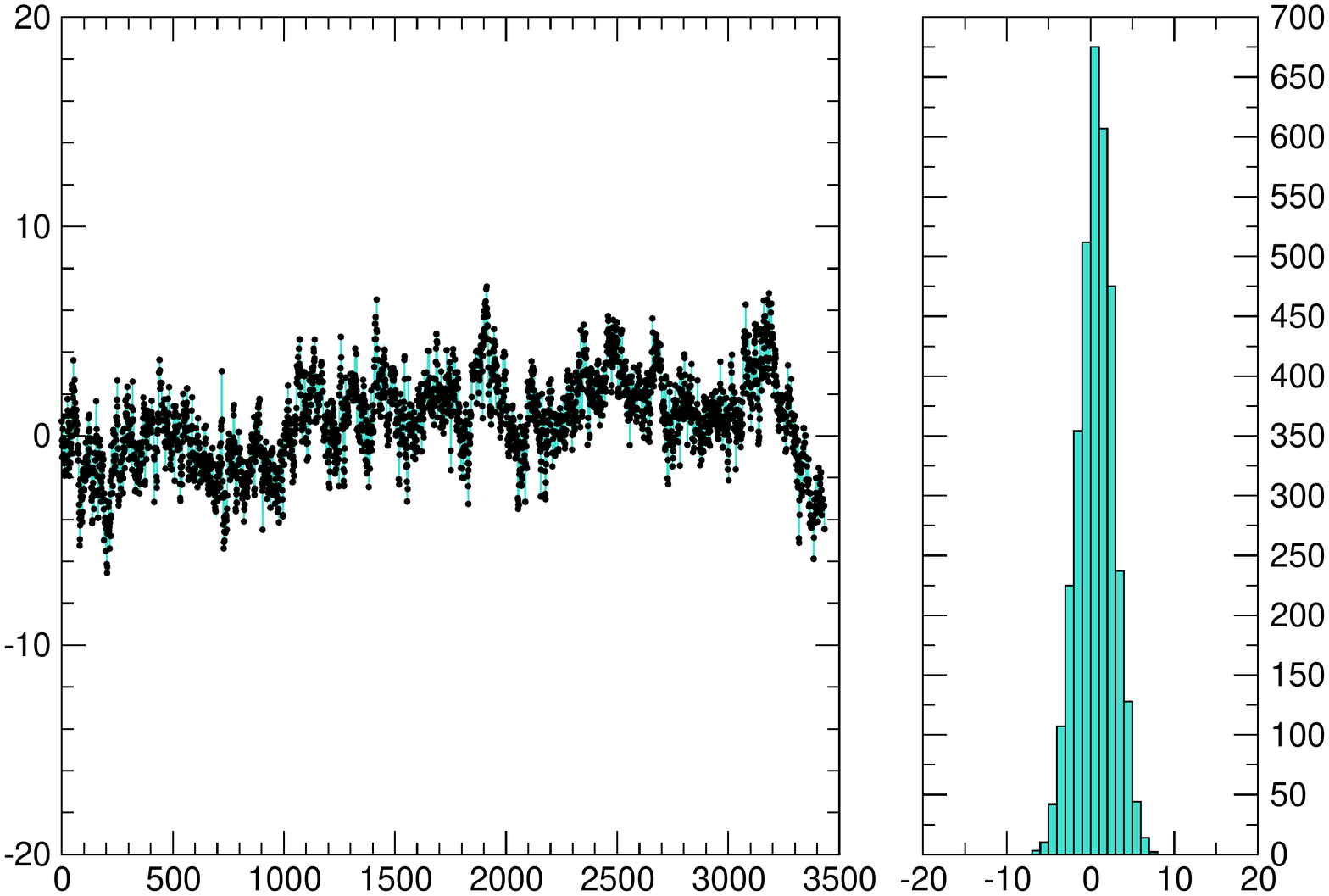} 
  \includegraphics*[scale=0.3]{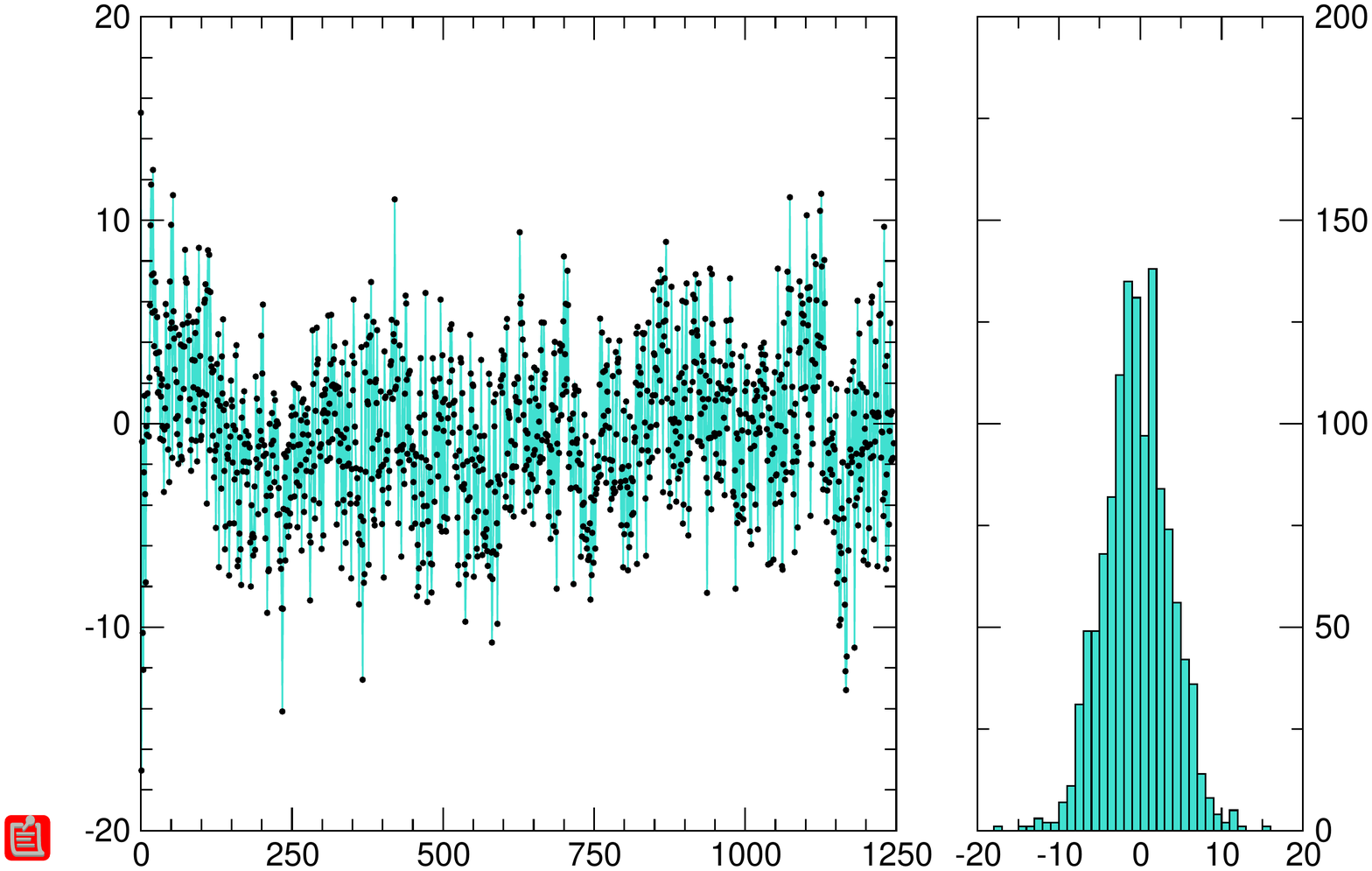}   
  \caption{Time history for the topological charge defined from the
    bosonic estimator in Eq.~\ref{eq:tTopCh} for the run G2, with
    periodic boundary conditions (left), and for the run OG1, with
    open boundary conditions (right). The topological charge is
    measured at flow time $t=2.0$.}
  \label{fig:PBCTC}
\end{figure}

\begin{figure}[h!]
  \centering
  \includegraphics*[scale=0.35]{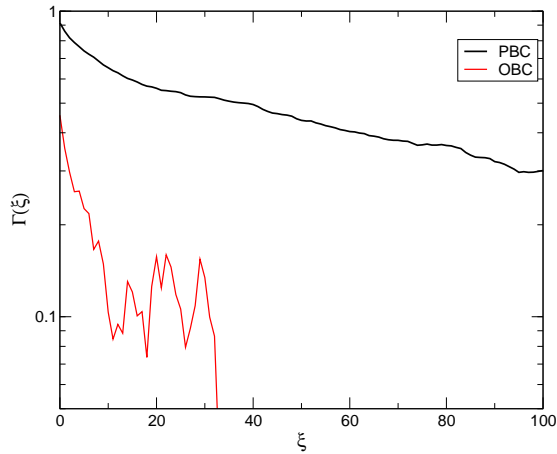}
  \caption{Normalized correlation function $\bar{\Gamma}(\xi)$ for the topological charge
    as a function of the time lag between measurements $\xi$. The
    black (respectively red) curve corresponds to the data obtained
    with periodic (respectively open) boundary conditions. }
  \label{fig:rhoplot}
\end{figure}

The results for the glueball spectrum are summarised in
Tab.~\ref{tab:gluetab}, where they are compared to the ones obtained
with periodic boundary conditions. 
\begin{table}[h!]
  \centering
  \begin{tabular}{|c|r|r|r|}
    \hline
     & $A^{++}$ & $E^{++}$ & $T^{++}$ \\
    \hline
    OBC & 0.179(28) & 0.212(34) & 0.283(27) \\
    PBC & 0.256(18) & 0.230(23) & 0.303(30) \\
   \hline
  \end{tabular}
  \caption{Results for the glueball spectrum at $am=-1.15$. We show
    the results obtained with periodic boundary conditions (PBC), and
    with open boundary conditions (OBC) averaged
    over $N_\mathrm{pts}=29$ time slices, as explained in the
    text. As discussed in Sect.~\ref{sec:lighter-fermion-mass} 
    the determination of the statistical error with PBC is most likely 
    underestimated because of long autocorrelation times. }
  \label{tab:gluetab}
\end{table}
The table shows a discrepancy for the mass of the $A^{++}$
glueball. We will argue that the error for the case with periodic
boundary conditions may be underestimated because of very long
autocorrelation times.  The expected ordering of the masses, with the
scalar glueball lighter than its spin-2 counterpart, is observed when
using open boundary conditions. The picture is much less clear in the
case of periodic boundary conditions: the $A^{++}$ state is heavier
than the $E^{++}$, although the values are compatible within
errors. For both choices of the boundary conditions, we do not observe
the expected degeneracy between the $E^{++}$ and $T^{++}$
channels. Note that this degeneracy is only expected in the continuum
limit; our simulations being performed at only one value of the
lattice spacing, we cannot disentangle finite-volume effects from the
$O(a)$ lattice artefacts. However, the fact that the $E^{++}$ and the
$T^{++}$ glueballs are degenerate at the heavier value of the mass,
suggests that the discrepancy for lighter fermions is due to finite
volume effects. In order to understand better the origin of these
different patterns, we have investigated the integrated
autocorrelation time of the topological charge in both settings, since
a poor sampling of the topological sectors could in principle lead to
larger systematic errors. We measured the topological charge using
a bosonic estimator computed along the Wilson flow:
\begin{equation}
  \label{eq:tTopCh}
  Q_t = -\frac{a^4}{32\pi^2} \sum_x \epsilon_{\mu\nu\rho\sigma}
  \mathrm{tr} \left[G_{t,\mu\nu}(x) G_{t,\rho\sigma}(x)\right]\, ,
\end{equation}
where $G_{t,\mu\nu}(x)$ is the clover term for the field strength tensor
on the lattice constructed from the gauge links at flow time $t$~\cite{Luscher:2011kk}.
The time histories for the topological charge defined from the gauge
fields at flow time $t=2.0$ in the runs with periodic and open boundary
conditions, G2 and OG1 respectively, are shown in Figs.~\ref{fig:PBCTC}. 
The two plots suggest that the topological
sampling is much more efficient in the simulations with open boundary
conditions. A more quantitative assessment of the tunneling between
topological sectors can be obtained by computing the integrated
autocorrelation time for both runs. We use the Madras-Sokal definition
of the autocorrelation time~\cite{Madras:1988ei}, implemented
according to Refs.~\cite{Wolff:2003sm,Luscher:2005rx}. As an illustration of its
typical behaviour, we report in Fig.~\ref{fig:rhoplot} the normalized
autocorrelation function for the topological charge computed at flow
time $t=2.0$, as a function of the time lag $\xi$ between measurements:
\begin{equation}
  \label{eq:GammaCorr}
  \bar{\Gamma}(\xi) = \frac{1}{N-\xi} \sum_{i=1}^{N-\xi} \left(Q_i - \bar{Q}\right)
  \left(Q_{i+\xi} - \bar{Q}\right)\, ,
\end{equation}
where $N$ is total number of measurements, $Q_i$ denotes the i-th
measurement of the topological charge $Q$, and $\bar{Q}$ its average.
The data show that the correlation decays much faster for the case of
open boundary conditions. The integrated autocorrelation time
$\tau_\mathrm{int}$ is defined as:
\begin{equation}
  \label{eq:tauintMS}
  \tau_\mathrm{int} =  \frac12 + \sum_{\xi=1}^{W}
  \bar{\Gamma}(\xi)/\bar{\Gamma}(0)\, ,
\end{equation}
where $W$ is the size of the Madras-Sokal window. The integrated
autocorrelation times, which are reported in Tab.~\ref{tab:tauintTC},
reflect the longer correlations in the case of the periodic boundary
conditions.
The current runs on the periodic lattices are not long enough to
determine with confidence the autocorrelation time. However, it is
clear from the numbers obtained for $\tau_\mathrm{int}$ using the
Madras-Sokal procedure, and from the plot of the normalized
correlation, that the autocorrelation is much longer in the case of
the periodic boundary conditions than it is with open boundary
conditions.
\begin{table}[h!]
  \centering
  \begin{tabular}{|c|c|r|}
    \hline
    $t$ & b.c. &$\tau_\mathrm{int}$ \\
    \hline
    2.0 & OBC & 6.0(2.0) \\
    2.0 & PBC & 80.6(51.2) \\
    5.0 & OBC & 7.8(2.7) \\
    4.4 & PBC &  87.4(54.7) \\
    \hline
  \end{tabular}
  \caption{Autocorrelation time of the topological charge; for each
    choice of boundary conditions, the
    autocorrelation is computed for two values of the flow time $t$. }
  \label{tab:tauintTC}
\end{table}

In the light of these results, it is tempting to associate the
unexplained features of the glueball spectrum to  the lack of
decorrelation between topological sectors in the case of simulations
with periodic boundary conditions. A more detailed investigation of
the correlation between topology and the glueball spectrum is beyond
the scope of this paper, and is postponed to future
investigations. Here we will simply conclude that the simulations with
periodic boundary conditions are affected by sytematic errors in the
gluebal spectrum that we cannot fully quantify currently. It is
reassuring to note that the limited data we have with open boundary
conditions yield results that are in line with the expected ones in
the continuum limit.  

\begin{figure}[ht] 
  \includegraphics*[scale=0.35]{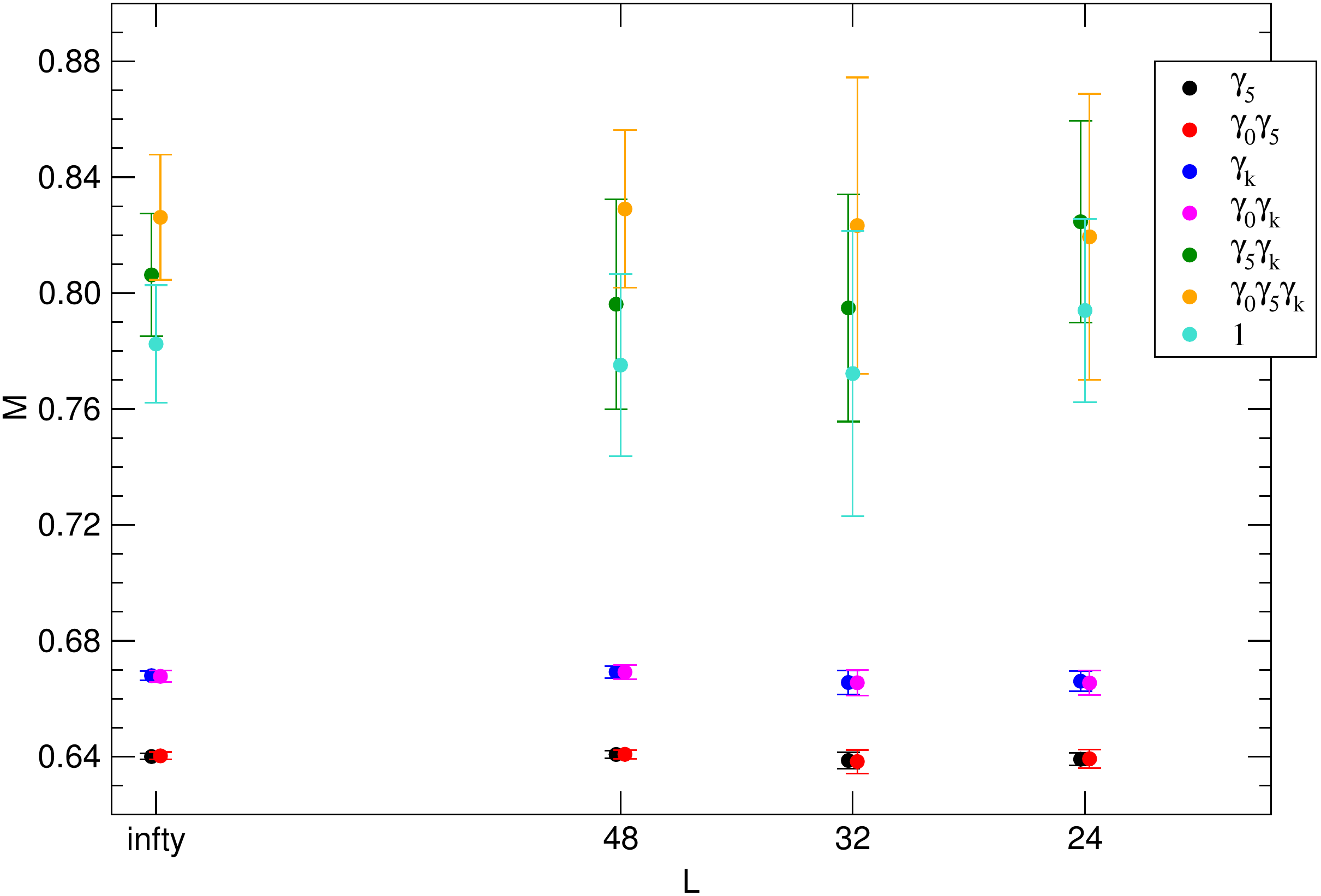} 
  \caption{ Summary of the spectrum of the mesonic states for the
    lattices included in this analysis at $am_0=-1.15$ for periodic
    boundary conditions. The data are shown as a function of the
    number of sites in the spatial directions $L/a$.  }
\label{fig:15mesons}
\end{figure}

\begin{figure}[ht] 
  \includegraphics*[scale=0.35]{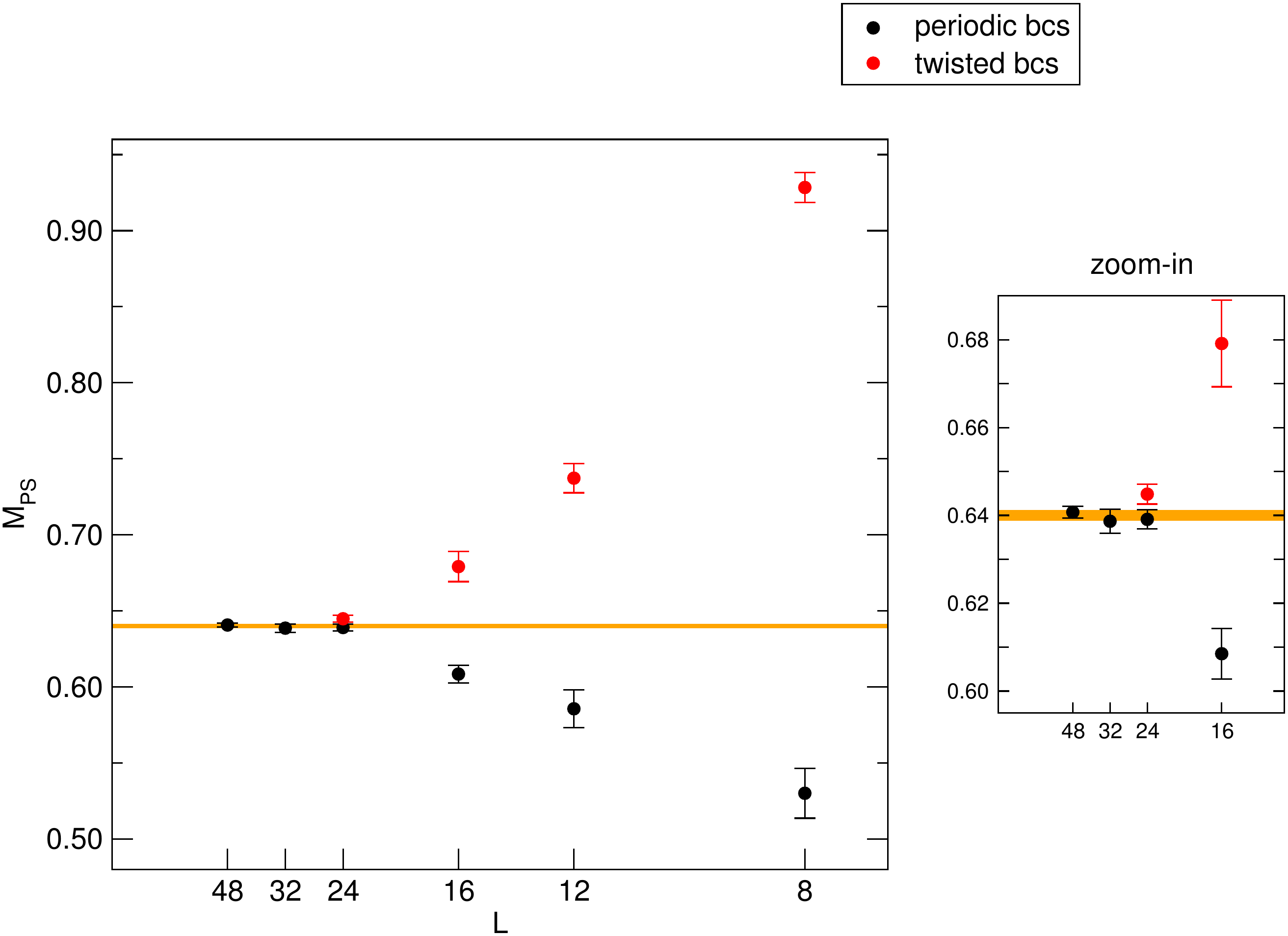} 
  \caption{
    Comparison of the pseudoscalar mass obtained from simulations with periodic and twisted boundary
    conditions respectively at $am_0=-1.15$. 
  }
\label{fig:15mps2}
\end{figure}

As already noted in previous studies, the meson spectrum is found to
reach its asymptotic value for smaller volumes than the glueballs. We
can therefore focus on a more quantitative analysis of the mesonic
spectrum at $am_0=-1.15$, which, for clarity, is reported on a larger
scale in Fig.~\ref{fig:15mesons}. The data show that the finite-volume
effects for this value of the mass are nicely under control, and that
a reliable infinite-volume limit can be taken.
In order to get a further estimate of the size of finite volume
effects, we compare the mass obtained on increasingly large volumes
using periodic and twisted boundary conditions. The results of these
comparison are reported in Fig.~\ref{fig:15mps2}, where independence
from the boundary conditions is seen for $L/a \geq 24$. We conclude
from this analysis that systematic errors due to finite volume effects
become smaller than 1\% for $L M_\mathrm{PS} \geq 15$.

\begin{figure}[ht] 
  \includegraphics*[scale=0.35]{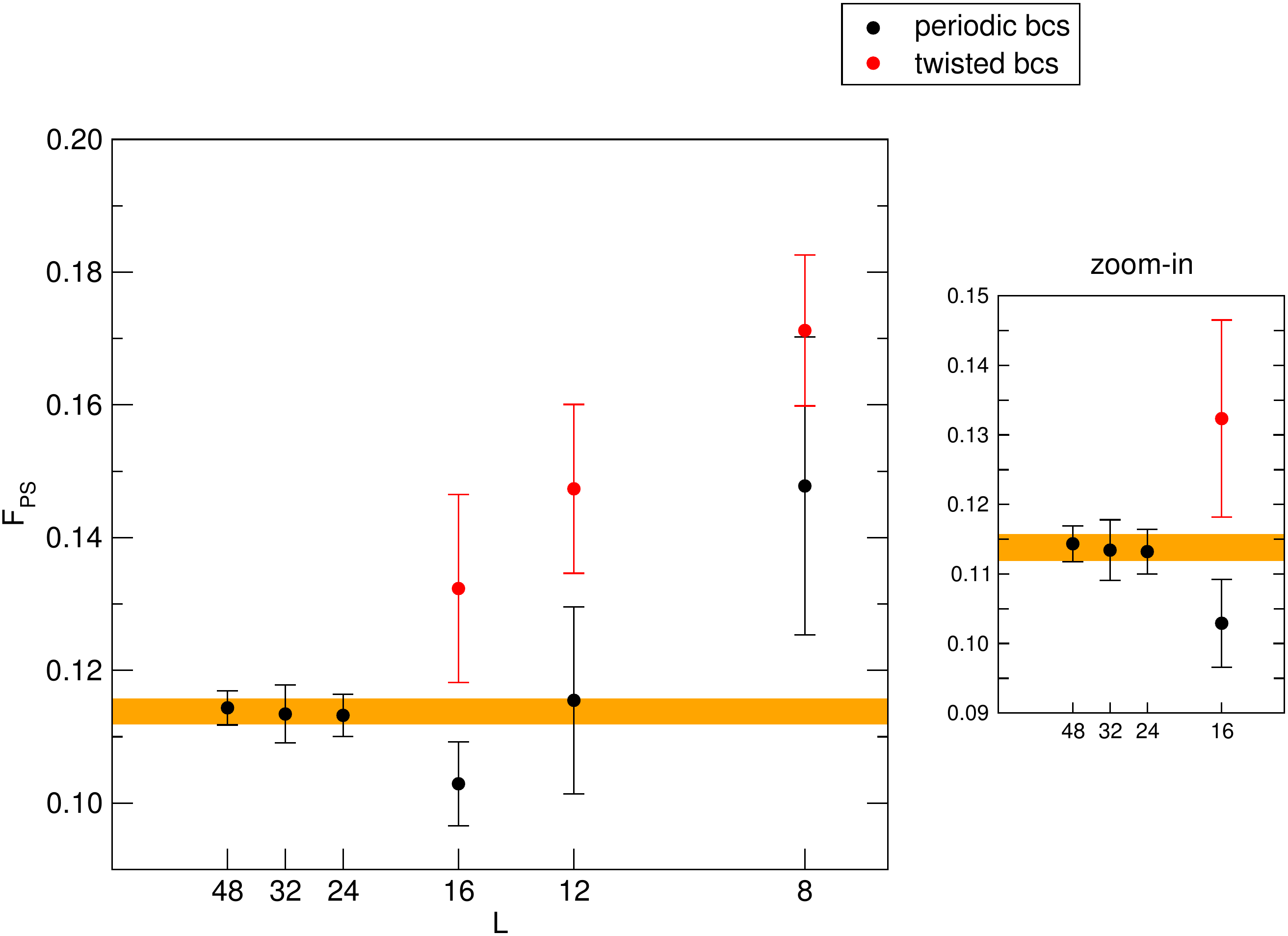} 
  \caption{
    Pseudoscalar decay constant as a function of the lattice size $L/a$ at $am_0=-1.15$. 
  }
\label{fig:15fps}
\end{figure}

Our previous studies had also highlighted a strong dependence of the pseudoscalar decay constant on
the size of the lattice. In order to get a more quantitative understanding of the problem, we have
measured $f_\mathrm{PS}$ on the larger lattices used for this analysis. The results are shown in
Fig.~\ref{fig:15fps}. The situation for the decay constant is more subtle: the behaviour is not
monotonic as a function of the lattice size, and we do not have a theoretical description of the
observed pattern. This unusual behaviour is due to the value obtained at $L/a=16$, which is only
2$\sigma$ away from the value obtained from the larger lattices. An extrapolation based on
the data for $L/a < 24$ would be affected by large systematic errors, when
compared to the value we obtain if only the larger volumes are
used. In particular the extrapolated value obtained from the smaller
lattices would end up being much smaller than the one extracted from the
large lattices, hence biasing the characterization of the
large-distance behaviour of the theory under study. It is worthwhile
emphasizing once again the importance of simulating near-conformal
theories in sufficiently large volumes. For each individual theory,
and for each choice of bare parameters in the action, the minimal
volume needs to be determined by careful numerical analyses. 

The data for the rest of the spectrum follow the pattern observed in
the examples above. For the mass range explored in this study, the
finite volume effects are below the 1\% level provided the spatial
size of the lattice is such that $L M_\mathrm{X}  \geq 5$. The mass
$M_\mathrm{X}$ that determines the mimimum size of the lattice is the
lightest mass in the spectrum. Note that for our lattices the lightest
state is always a glueball state. As a rule of thumb, we can translate the bound
above into a constraint for the mass of the pion, which is approximately
three times heavier than the glueball; the condition to avoid
finite-volume effects becomes $L M_\mathrm{PS} \geq 15$.

In the regime where finite-volume effects are below 1\%, we can study
the ratio of masses of different states in the spectrum. A consequence
of the existence of an IRFP is that all states in the spectrum have to
scale to zero as a function of the fermion mass, at a common rate
dictated by the mass anomalous dimension $\gamma_*$:  
\begin{equation}
  \label{eq:MHscal}
  M_X \propto m^{1/(1+\gamma_*)}\, ,
\end{equation}
where $X$ denotes a generic state in the spectrum of the theory. 
This simple scaling laws implies that the spectrum of a theory with an
IRFP is different from the spectrum of QCD already at a qualitative
level. The lack of spontaneous symmetry breaking in a conformal
theory yields the absence of light Goldstone bosons, and hence the
ratio of the Goldstone boson mass to the mass of other states in the
spectrum does not vanish. 

\begin{figure}[ht] 
  \includegraphics*[scale=0.35]{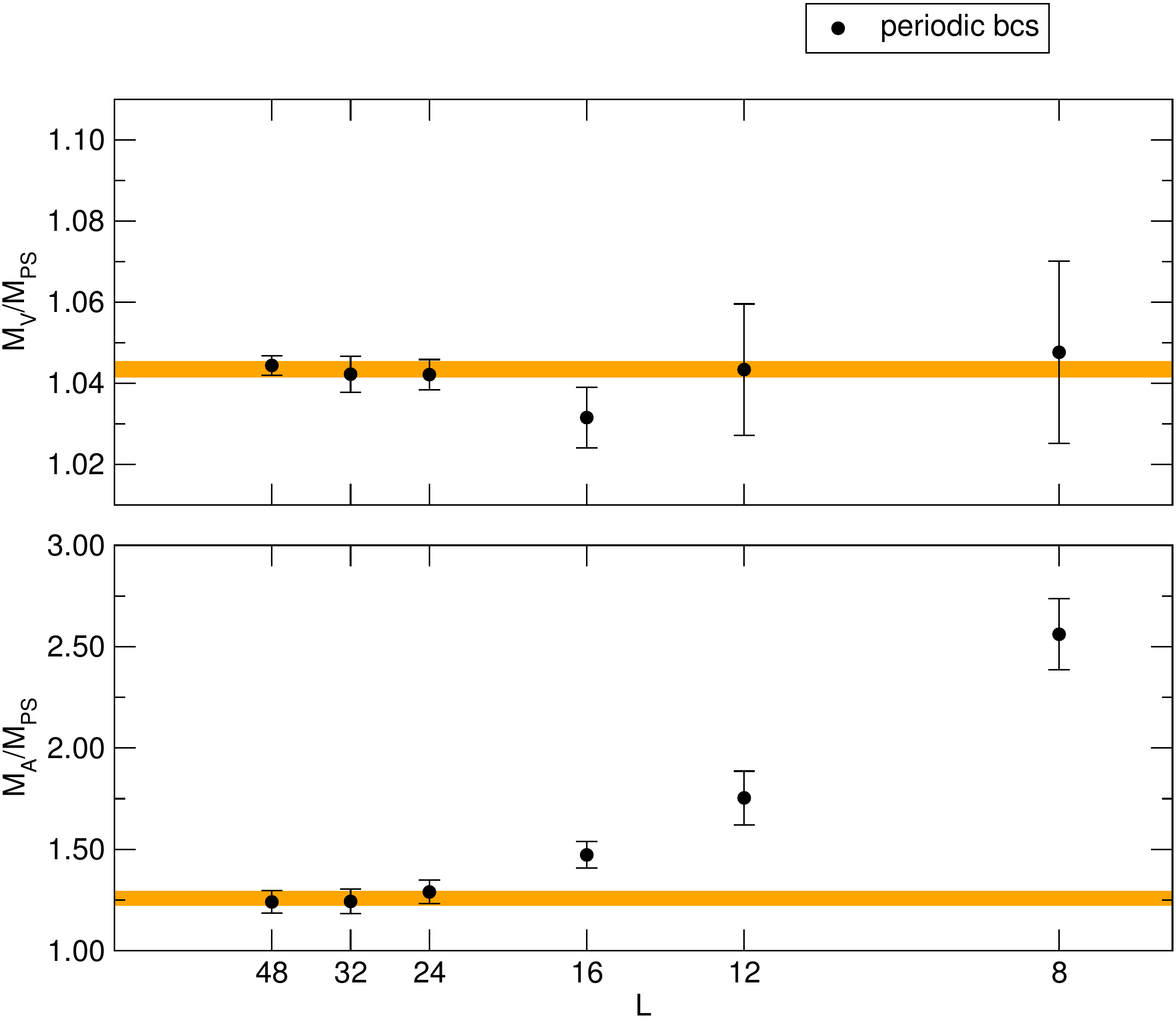} 
  \caption{
    Ratios of the mass of the vector state (upper plot) and axial
    vector state (lower plot) to the mass of the pseudoscalar
    state. The data are obtained at the lightest mass, and plotted as
    a function of the lattice volume $L$. 
  }
\label{fig:15ratios}
\end{figure}

The data presented in Ref.~\cite{DelDebbio:2010hx} suggest that the mass of the
pseudoscalar state scales indeed at the same rate as the vector mass for SU(2)
with adjoint fermions, thereby going to a constant in the zero-mass
limit. However some dependence on the finite size of the lattice remains
visible for the smallest lattices. The recent runs on larger volumes
presented in this paper have settled this issue for the two masses that we have
considered. Fig.~\ref{fig:15ratios} shows the ratio of the mass of the vector
and axial vector states to the mass of the pseudoscalar state for the lightest
mass simulated. The results for the ratio become independent of
the lattice size in both channels for $L>24$.

\subsection{Extrapolated results}
\label{sec:extrapolated-results}

The data from the larger lattices can be extrapolated to the
thermodynamical limit. We used a fit to a constant including the data
from lattices with linear size $L\geq 24$ for the light mass, and $L
\geq 16$ for the heavy one. The results of these fits
are already reported in the figures presented above. We summarise the
results for the extrapolated values for both values of the fermion
mass in Tab.~\ref{tab:thermsum}. All values are given in units of the
lattice spacing. 
\begin{table}[h!]
  \centering
  \begin{tabular}{|c|r|r|r|r|r|r|r|r|r|r|r|}
    \hline
    $-am_0$&$\sqrt\sigma$&$A^{++}$&$E^{++}$&$T^{++}$&$\gamma_5$&$\gamma_0\gamma_5$&$\gamma_0\gamma_k$&$\gamma_k$&$1$&$\gamma_5\gamma_k$&$\gamma_0\gamma_5\gamma_k$
    \\
    \hline
    1.05 & 0.156(1) & 0.542(17) & 0.77(3) & 0.80(4) & 1.183(1) &
                                                                 1.183(1)&
                                                                               1.232(2)&
                                                                                                  1.232(1)&
                                                                                                            1.583(42)&
                                                                                                                       1.593(53)&
                                                                                                                                  1.595(80)
    \\
    \hline
    1.15 & 0.0614(8) & --- & --- & --- & 0.6401(11) & 0.06402(13)&0.6677(19)&
                                                            0.6680(16)&
                                                                        0.7824(20)&
                                                                                    0.806(21)&
                                                                                               0.826(22)
    \\
    \hline
  \end{tabular}
  \caption{Summary of the infinite volume spectrum of the theory for
    both values of the fermion mass. The values reported in the table are obtained by
    extrapolating the results from our largest lattices. Given
    the systematic errors that are still visible in the glueball
    spectrum, we do not provide extrapolated values for the mass of
    the $E^{++}$ and the $T^{++}$ states for the lighter fermion
    mass. All values are given in units of the lattice spacing. }
  \label{tab:thermsum}
\end{table}
It is interesting to point out that the degeneracies in the spectrum
follow the expected pattern, both for the glueball sector and the
mesonic sector. This suggests that the lattice artefacts are
relatively small in our simulations, and the discrepancies from the
expected spectrum vanish as long as we can take the thermodynamical
limit. In some channels, like e.g. the axial vector, the discrepancy
between the result at $L/a=16$ and the extrapolated result at infinite
volume is as large as 16\%.

\begin{figure}[h!]
  \centering
    \includegraphics*[scale=0.35]{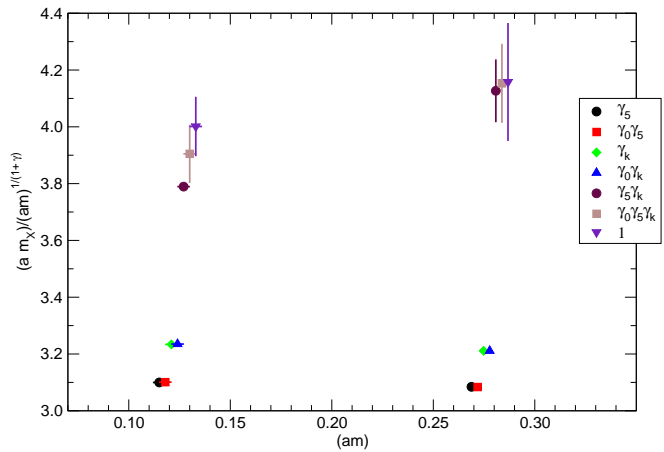} 
  \caption{Scaling behaviour of the mesonic spectrum. The mass of the
    state $X$ is denoted by $M_X$, while $m$ denotes the bare mass of
    the fermionic field. The rescaled
    data become independent of the value of the fermion mass,
    which is in agreement with the expected behaviour in the presence
    of an IRFP. Different colors correspond to different quantum
    numbers of the mesons. Some points are shifted along the
    horizontal axis for clarity. Note that the scaling exponent is not
    determined by a fit to the data; its value is taken
    from existing studies in the literature, as explained in the text.}
  \label{fig:therm-scal}
\end{figure}

The scaling dictated by the existence of the IRFP, see
e.g. Eq.~(\ref{eq:MHscal}), should be obeyed by the extrapolated
spectrum. Fig.~\ref{fig:therm-scal} shows that the scaling ratio
\begin{equation}
  \label{eq:therm-scal}
  (aM)/(am)^{1/(1+\gamma_*)} 
\end{equation}
remains constant for both values of the fermion mass, as expected from
the scaling hypothesis. Note that the value of the critical exponent
$\gamma_*=0.371$ is taken from Ref.~\cite{Patella:2012da}, and is not
adjusted to the results of the current simulations.

\subsection{Finite-size scaling}
\label{sec:finite-size-scaling}

In the presence of an IR fixed point, the dependence of the spectrum and
matrix elements on the fermion mass and volume are determined by
finite-size scaling (FSS). For sufficiently small fermion masses and
sufficiently large volumes the dependence of field correlators is
determined by scaling relations that are obtained from the linearized
RG equations in a neighbourhood of the fixed point. The scaling laws
for the theory under consideration here were studied in detail in
Refs.~\cite{DeGrand:2009mt,DelDebbio:2010hx,DelDebbio:2010ze,DelDebbio:2010jy,DelDebbio:2013qta}.
Here we simply recall the results for the masses in the spectrum:
\begin{equation}
  \label{eq:FSSscal}
  L M_X = f_X(x)\, ,
\end{equation}
where $M_X$ is the mass of the physical state $X$, $x=L^{y_m} m$ is the
so-called scaling variable, and $y_m = 1+\gamma_*$. The function $f_X$
is a universal function, in the sense that data generated at different
values of the gauge coupling, fermion mass, and volume should all be
described by the same function. The only theoretical constraint on
$f_X$ comes from the fact that it should reproduce the infinite-volume
scaling with the mass of the fermion as $L$ becomes large. It is easy
to verify that:
\begin{equation}
  \label{eq:FSSuniv}
  f_X(x) \stackrel{x\to\infty}{\sim} A_X x^{1/y_m}
\end{equation}
satisfies this condition. 

\begin{figure}[ht] 
  \includegraphics*[scale=0.35]{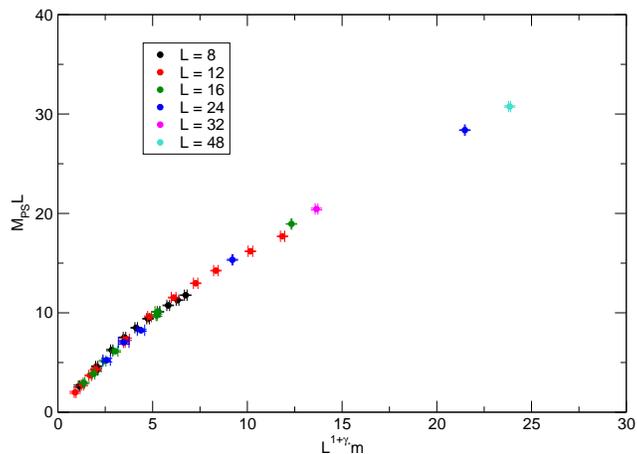} 
  \caption{
    Finite-size scaling of the pseudoscalar mass. The data obtained on
    different volumes combined using the scaling exponent determined
    in Ref.~\cite{Patella:2012da} fall on a universal curve as expected. 
  }
\label{fig:mpsFSS}
\end{figure}

\begin{figure}[ht] 
  \includegraphics*[scale=0.35]{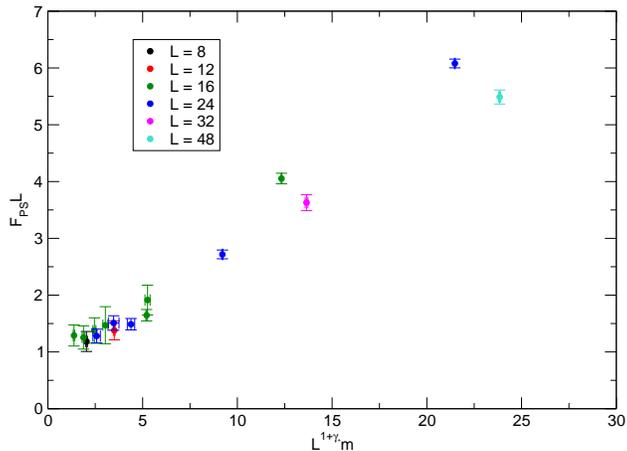} 
  \caption{
    Finite-size scaling of the pseudoscalar decay constant. The data obtained on
    different volumes combined using the scaling exponent determined
    in Ref.~\cite{Patella:2012da} fall on a universal curve; however the
    scaling deviations for the decay constant are larger than the ones observed
    for the pseudoscalar mass. 
  }
\label{fig:fpsFSS}
\end{figure}

For each observable we expect a different function, but then that
single function must describe the behaviour of data from lattices with
different volumes. The data shown in Figs.~\ref{fig:mpsFSS}
and~\ref{fig:fpsFSS} are in excellent agreement with the scaling
dictated by the renormalization group~\footnote{It is worth recalling
  that scaling is expected in the limits of large size and small
  mass. Observable-dependent deviations from scaling should be
  expected at small volumes or large masses.}, thereby strenghtening
the hypothesis that this theory does indeed possess a fixed point that
governs the large-distance dynamics. The scaling of the variables with
volumes is obtained using the critical exponent $\gamma_*$ determined
in Ref.~\cite{Patella:2012da} from the spectrum of the Dirac
operator. Once again note that $\gamma_*$ is not a free parameter in
this study; it is instead fixed to the best value available in the
literature.

\section{Conclusions}
\label{sec:conclusions}

In this paper we have reported the results of our latest large-volume
runs for the spectrum of SU(2) gauge theory with two adjoint
fermions. This study was motivated by the finite-volume effects seen
in previous simulations, which were potentially obscuring the scaling
expected for a theory with an IR fixed point of the RG flow.

The data show that large volumes are mandatory for a robust
quantitative determination. We have shown that smaller volumes are
enough to control finite-size effects on the triplet meson masses,
however much larger volumes are needed in order to have a good control
over glueball finite-size effects. This is not surpising considering
that this theory shows a separation of scales between the two sectors.

Finite-volume effects on masses of triplet mesons drop below $5\% $ if
$L M_\mathrm{PS} \simeq 9.5$. We have substantiated this statement with
a systematic comparison of triplet meson masses calculated for
different values of the spatial volume, with periodic and twisted
boundary conditions. In the gluonic sector our conclusions are less
definite, because of the intrinsic difficulty in measuring glueball
masses. As a rule of thumb, a spatial size such that
$L \sigma^{1/2} \gtrsim 3$ is desirable in order to avoid large
finite-volume effects in the glonic sector.

We have analysed in detail two values of the fermion mass, which
produce compatible ratios of triplet-meson masses. These findings
strengthen our previous conclusions on the existence of a fixed point
for this theory. The fermion masses used in this study seem to be
light enough in order for the triplet meson 
spectrum to be in the scaling region of the IR fixed point.
Moreover the data obtained on different volumes fall on a universal
curve using the value of the mass anomalous dimension previously
determined in Ref.~\cite{Patella:2012da}. On the other hand our
infinite-volume estimate for the ratio $M_\mathrm{PS}/\sqrt\sigma$ changes
substantially with the fermion mass, suggesting that lighter masses
are necessary in order to access the scaling region of the gluonic
spectrum.

Should such theories become serious candidates for
phenomenology, we have presented convincing evidence that physically relevant
informations on their spectrum can only be extracted from large
volume studies.

\section*{Acknowledgments}
LDD kindly acknowledges the hospitality of the Theory Department at
CERN, where part of this work has been carried on. LDD is supported by
STFC, grant ST/L000458/1, and the Royal Society, Wolfson Research
Merit Award, grant WM140078.  BL is supported by STFC (grant
ST/L000369/1). CP is supported by a
Lundbeck Foundation fellowship and by the CP$^3$-Origins centre which
is partially funded by the Danish National Research Foundation, grant
number DNRF90. AR is supported by the Leverhulme Trust, grant
RPG-2014-118, and STFC, grant ST/L000350/1.
The numerical computations have been carried out using resources from the DiRAC Blue Gene Q Shared Petaflop system, HPC Wales and the High Performance Computing Center (HPCC) Plymouth. 
The DiRAC Blue Gene Q Shared Petaflop system is operated by the Edinburgh Parallel Computing Centre on behalf of the STFC DiRAC HPC Facility (www.dirac.ac.uk). This equipment was funded by BIS National E-infrastructure capital grant ST/K000411/1, STFC capital grant ST/H008845/1, and STFC DiRAC Operations grants ST/K005804/1 and ST/K005790/1. DiRAC is part of the National E-Infrastructure. 
The HPC Wales infrastructure is supported by the ERDF through the WEFO, which is part of the Welsh Government.

\bibliographystyle{apsrev}

%\bibliography{/Users/ldd/sync/hirep/doc/bib/hirep,/Users/ldd/sync/hirep/doc/bib/myhirep,/Users/ldd/sync/hirep/doc/bib/QFT,/Users/ldd/sync/hirep/doc/bib/lat,/Users/ldd/sync/hirep/doc/bib/pheno,other}
\end{document}